\documentclass[12pt]{article}
\usepackage[pctex32]{graphics}
\begin{document}
\vspace{2cm}
\begin{center}
{\bf  \Large  Finite-Temperature Cosmological Phase Transition in a
Rotating Spacetime}
\vspace{1cm}

                      Wung-Hong Huang\\
                       Department of Physics\\
                       National Cheng Kung University\\
                       Tainan,70101,Taiwan\\

\end{center}
\vspace{2cm}
 We use the $\zeta$-function regularization method to evaluate the finite temperature 1-loop effective potential for  $\phi^4$ theory in the Godel spacetime. It is used to study the effects of temperature and curvature coupling on the cosmological phase transition in the rotational spacetime. From our results the critical temperature of symmetry restoration, which is a function of curvature coupling and magnitude of spacetime rotation, can be determined.

\vspace{3cm}
\begin{flushleft}
E-mail:  whhwung@mail.ncku.edu.tw\\
Classical and Quantum Gravity Vol.8 (1991) 1471-1479
\end{flushleft}


\newpage
\section{Introduction}

    The inflationary Universe model proposed by Guth [1] and its improved version by Linde [2], Albrecht and Steinhardt [3], is based on the theory of phase transition in the grand unified theories [4]. According to Guth's scenario,, as the Universe expands and cools down, a phase transition occurs from the symmetric state to the asymmetric state. During the phase transition the energy of the Universe will be dominated by the vacuum energy of the Higgs field, and the Universe is in a de Sitter state. It is this exponential expansion that the Universe has with a scale factor of at least 28 orders of magnitude, which resolves many long-standing cosmological puzzles such as the horizon and flatness problems. However, as was noted by Guth himself, the original inflationary scenario would result in an extremely large inhomogeneous and anisotropic Universe.

    This problem can be avoided in the improved version [2, 3] in which the first-order phase transition in Guth's model is replaced by the Coleman-Weinberg [5] symmetry breaking. Because the Coleman-Weinberg potential is very flat near the origin, the Higgs field will roll down the effective potential very slowly and the Universe keeps expanding exponentially. It is during.this period of slow growth that a single bubble inflates sufficiently to cover the entire Universe which will then result in a very homogeneous and isotropic Universe as we observe now.
    The success of the improved inflationary Universe model has led many authors to study quantum fields in curved space [ 6] to see the gravitational effect on the cosmological phase transition [7, 8]. From the effective potential they calculate, it was found that the scalar gravitational coupling $\zeta$  and the magnitude of the scalar curvature R crucially determine the fate of symmetry. One knows that, in the inflationary scenario, the isotropy of the Universe before the phase transition is an a priori assumption. Although the quantum field effects on the curved space could account for this, such as the isotropization by particle production [9], it is less clear whether an initially large anisotropy can indeed be smoothed out at a sufficiently early stage, Therefore it is interesting to investigate the anisotropic effect [10-12] on the Coleman-Weinberg symmetry breaking. For the same reason, it is also worthwhile to investigate the effects of inhomogeneity [13! and rotation [ 14] on the cosmological phase transition.

    In this paper we will analyze the process of Coleman-Weinberg symmetry breaking in a rotational spacetime of Godel metric [15] at finite temperature. The zero-temperature theory in Godel spacetime has been studied in [14]. Note that the
finite-temperature field theory in flat spacetime was investigated several tears ago [16].   It has also been studied in the static [17] and dynamic spacetimes [18]. (A more complete list of references on the finite-temperature field theory in curved spacetimes can be found in [18].)

    We know that the problem of obtaining a solution to Einstein's equation for the
rotating spacetime is fairly old. Godel [15] was the first to find an exact rotating
solution. However, this solution suffers numerous pathologies, such as the violation of causality, the absence of a global Cauchy surface, the non-orientability of time, etc (see the discussion in [19] for details). This solution also has no expansion and no shear. Although the Godel model is not very physical, we study it here because it appears as the simplest rotating cosmological model. Also, despite the fact that the cosmological rotation is very small by the present observations it cannot be completely ruled out, at least in the early Universe.

   The paper is organized as follows. We begin in section 2 by solving the scale wave equation in the Godel spacetime and setting a formula to obtain the effective potential which, however, is formally divergent. In section 3 we use the $\zeta$ function regularization method [20] to render it finite. A short discussion and conclusion are given in the last section.

\section{ Formalism}

The action describing a massive (m), self-interacting ($\lambda$) scalar field ($\phi$) coupled arbitrarily ($xi$) to the gravitational background is

$$ S[\tilde \phi] = \int d^4x \sqrt {-g} \left( {1\over 2}g^{\mu\nu}\partial_\mu \tilde \phi \partial_\nu \tilde \phi - {m^2\over 2}\tilde \phi ^2 -{\xi \over 2}R\tilde \phi ^2 - {\lambda\over 4!}\tilde \phi ^4  \right).    \eqno{2.1)}$$
\\
The effective potential $V_{eff}$ is given by [6]

  $$exp \left ( -i \int d^4x \sqrt {-g} V_{eff} \right) = \int D\tilde \phi e^{iS[\tilde \phi ]} = det^{-1/2} [\triangle + M^2],    \eqno{(2.2)}$$
\\                                                                                                                                                                                                                                                             where $\triangle \equiv g^{\mu\nu}\nabla _\mu \nabla_\nu$ and $M^2 = m^2+\xi R+{1\over2}\lambda \phi^2$, with $\phi=<\tilde\phi>$.  The above equation can be rewritten as

$$\int d^4x \sqrt {-g} V_{eff} = -{1\over2}i Tr ln(\triangle + M^2). \eqno{(2.3)}$$
\\
The trace can be easily evaluated once we know the eigenvalue of the operator ($\triangle + M^2$).  Denoting the eigenvalue by $\eta$ and eigenfunction by h, then the theory in the Godel spacetime with a metric form

$$d^2 s = (dt +e^{\alpha x} dy)^2 -dx^2 -{1\over2} e^{2\alpha x} dy^2-dz^2, \eqno{(2.4)}$$
\\
we have the equation

$$h_{,tt} - 4 e^{-\alpha x}h_{,ty}  + 2 e^{-2\alpha x}h_{,yy} + h_{,xx}+\alpha h_{,x}  +h_{,zz}  - M^2 h + \eta h = 0.    \eqno{(2.5)}$$
\\
Assuming the following form for the solution

$$h( t, x)=exp(ik_y y+ik_z z -i\omega t) \Psi (x)  \eqno{(2.6)}$$
\\
we can find from (2.5) find the differential equation of  $\Psi (x)$

$$\Psi_{,tt} + \alpha \Psi_{,x} - [\omega^2 + k_z^2 + 2 k^2_y e^{-2\alpha x} + 4  k^2_y  \omega e^{-\alpha x} + M^2 - \eta ] \Psi = 0.      \eqno{(2.7)}$$
If we introduce a new variable
$$  u = \alpha^{-1} 2 \sqrt 2 |k_y| e^{-\alpha x}, \eqno{(2.8)}$$
for the case of  $k\ne 0$, then (2.7) becomes

$$ \Psi_{,uu} - \left({1\over 4}+{\sqrt 2 \omega \epsilon\over \alpha u} +{\omega^2+k_z^2+M^2-\eta\over \alpha^2 u^2} \right) \Psi = 0,\eqno{(2.9)}$$
\\
with $\epsilon  = k_y/|k_y|$, which is the Whittaker's equation and its solutions are well known.   From the asymptotic behavior of the solutions one can see that the condition to have an everywhere-bounded solution is

     $$\eta = k_z^2 + M^2 +\alpha^2[n+{1\over2}] +{1\over4}\alpha^2 - [\omega + \sqrt 2 \epsilon \alpha (n+{1\over2})]^2      \eqno{(2.10)}$$
\\
with n =0, 1, 2,  ....  Note that the scalar wave equation in the Godel Universe has
been solved in [21] and [22], where the quantized dispersion relation (Let $\eta$=0 in (2.10)) was first derived. 

    When $k_y=0$, we can define a new variable
        $$ v= e^{-\alpha x},   \eqno{(2.11)}$$
then (2.7) becomes

$$ \Psi_{,uu} - {\omega^2+k_z^2 \over \alpha^2 v^2}~ \Psi = 0,  \eqno{(2.12)}$$
\\
The solutions are of simple powers of ,v and not bounded everywhere; thus we neglect it.

    To proceed, let us write

$$ Tr ln(\triangle + M^2)  = \int d^4 x \sqrt {-g} <t,x|ln(\triangle + M^2)|t,x >~~~~~~~~~~~~~~~~~~$$
$$  = \int d^4 x \sqrt {-g} \sum_K <t,x|K>^2 <K|ln(\triangle + M^2)|K>,  \eqno{(2.13)}$$
where

$$ <t,x|K> = (2\pi)^{-3/2} \Psi_L\left({1\over\alpha} 2 \sqrt 2 |k_y| e^{-\alpha x}\right) exp (ik_y y +ik_z z -i\omega t),  \eqno{(2.14)}$$
$$\sum_K= (-g)^{-1\over2}\int d\omega \int dk_y \int dk_z \sum_n ,  \eqno{(2.15)}$$
\\
with K and L denote the sets of eigenvalues $(\omega, k_z, k_y, n)$ and $(\omega, k_z, n)$ respectively.

   Two remarks are in order. (a) We need not worry about the existence of an
integration over $k_y$ in (2.15) despite the fact that the eigenvalue is independent of $k_y$.

    In fact, this integration over $k_y$,, could be used to perform the integration over the argument of' $\Psi$ and thus makes the calculation of (2.13) possible, i.e.

$$ \sum_K <t,x|K>^2 <K|ln(\triangle + M^2)|K > \hspace{8cm}$$
$$ = (-g)^{-1\over2}\int d\omega \int dk_z~ \sum_n  <\omega, k_z, n|ln(\triangle +M^2)|\omega, k_z, n>$$
$$\times (2\pi)^{-3}\int dk_y |\Psi_L\left({1\over\alpha} 2 \sqrt 2 |k_y| e^{-\alpha x}\right)|^2 $$
$$= (2\pi)^{-3}\int d\omega \int dk_z \sum_n  <\omega, k_z, n|ln(\triangle +M^2)|\omega, k_z, n> \eqno{(2.16)}$$ 
\\
if the function $\Psi$  is normalized as $\int du |\Psi(u)|^2= 1$. (b) The weight $(-g)^{-l/2}$ in (2.15) should be introduced because $d\omega  dk$ is a tensor density rather than a tensor. Notice that we introduce a factor $(-g)^{1\over2}$ for $dx$,     but $(-g)^{-1\over2}$ for $d\omega dk_z$. This is because the weight of the density for a contravariant vector is the inverse of that for a covariant vector [23]. (The function $exp(ik_y+ik_z-i\omega t)$  in (2.6) tells us that $(t,x)$ and $(\omega , k)$ are the contravariant vector and the covariant vector respectively.) It is the introduced factor $(-g)^{-1\over2}$ which will cancel the factor $e^{\alpha x}$ coming from the change of the variable from $k_y$, to u, we can therefore have an x-independent effective potential. (Note that the Godel Universe is homogeneous in both time and space.) It seems that several previous articles have neglected this point.

    We therefore obtain
$$ V_{eff} = {T\alpha\over 8\pi^2} \int dk_z\sum_n\sum_l <k_z, n, l|ln(\triangle +M^2)|k_z, n,l>  \eqno{(2.17)}$$   
\\
where we have used the following replacement

$$\omega + \sqrt 2 \epsilon \alpha (n+{1\over2}) \rightarrow  i2\pi lT, \eqno{(2.18)}$$ 
in order to perform a computation at finite temperature T. The expression (2.17) is divergent and needs to be regularized This is the work of section 3.

\section {The $\zeta$-function regularization}

We now use the method of  $\zeta$ function regularization [20] to find the renormalized value of $V_{eff}$ in (2.17). Introducing a complex parameter  $\nu$, we obtain a relation

$$ \int dk_z \sum_n \sum_l  <k_z, n,l |ln(\triangle +M^2)|k_z, n,l> = -{\partial\over \partial \nu} \zeta (\nu)|_{\nu=0},        \eqno{(3.1)}$$
\\
where
 $$ \zeta (v) =  \int dk_z \sum_n \sum_l  <k_z, n,l | (\triangle +M^2)|k_z, n,l>^{-\nu}.    \eqno{(3.2)}$$
\\
Using (2.10) and (2.18) we have
$$ \zeta(\nu) = \int dk_z \sum_n\sum_l [k^2_z+M^2+\alpha^2(n+{1\over2})^2+{1\over 4}\alpha^2+4\pi^2 l^2T^2]^{-\nu}$$
$$ = {\Gamma (1/2)\Gamma (\nu-1/2)\over \Gamma (\nu)} \sum_n \sum_l  [M^2+\alpha^2(n+{1\over2})^2+{1\over 4}\alpha^2+4\pi^2 l^2T^2]^{-\nu}$$
$$ =  {\Gamma (1/2)\Gamma (\nu-1/2)\over \Gamma (\nu)} (\sum_n \sum_l  \left(M^2+{1\over 4} \alpha^2 n^2+{1\over 4}\alpha^2+4\pi^2 l^2T^2\right)^{1/2-\nu}$$
$$ - \sum_n \sum_l  \left( M^2+\alpha^2 n^2+{1\over 4}\alpha^2+4\pi^2 l^2T^2\right)^{1/2-\nu}).   \eqno{(3.3)}$$
\\
Although the above summation is formally divergent, it could be well defined through the analytic continuation with the formula [24, 25]

$$\sum_{n,l} \left(M^2 + (n/a)^2  + (l/b)^2\right) ^{-\nu} = ab(M^2)^{1-\nu} \pi \Gamma (\nu ) \left( \Gamma (\nu-1) +2 \sum_{n,l} {}^{'} Z^{\nu-1} K_{1-\nu}(2Z) \right), \eqno{(3.4)}$$
\\
where K is the modified Bessel function and Z is defined by
               $$ Z = \pi M[ n^2 a^2 + 1^2 b^2]^{1/2}.                     \eqno{(3.5)}$$ 
A prime on the summation denotes that the term $n = l = 0$ is omitted. Using this formula (3.3) becomes

$$ \zeta(\nu) = (2\pi \alpha T)^{-1} {1\over \Gamma (\nu)} \pi^\nu [\pi ({1\over4}\alpha^2+M^2)]^{3/2-\nu} \hspace{5cm}$$
$$\times \left(-{1\over2} \Gamma(\nu-3/2) + \sum_{n,l} {}^{'}[2Z_1^{\nu-3/2} K_{3/2-\nu}(2Z) - Z_2^{\nu-3/2} K_{3/2-\nu}(2Z) \right), \eqno{(3.6)}$$
where
$$ Z_1 = \pi\left( ({1\over4}\alpha^2+M^2)(l^2/4\pi^2 T^2 + 4n^2/\alpha^2)\right)^{1/2}$$
$$ Z_2 = \pi\left(({1\over4}\alpha^2+M^2)(l^2/4\pi^2 T^2 + n^2/\alpha^2)\right)^{1/2}\eqno{(3.7)}$$
\\
As $\Gamma(\nu) \rightarrow  1/\nu$  we obtain from (2.17) the final result

$$V_{eff} = (16\pi^3)^{-1} [\pi ({1\over 4} \alpha^2 + M^2)]^{3/2} \left( {2\over 3}\sqrt \pi - \sum_{n,l}{}^{'}[2Z_1^{\nu-3/2} K_{3/2-\nu}(2Z) - Z_2^{\nu-3/2} K_{3/2-\nu}(2Z)] \right). \eqno{(3.8)}$$

   A comment should be mentioned before applying the above exact formula to
analyze the cosmological phase transition. From the formula of analytic continuation (3.4) one sees that the cases of $\alpha  = 0$  and/or $T=0$  will not make sense. However, (3.8) will give $V_{eff}  \rightarrow (1/24\pi)M^3$,  as each term in the summation will approach zero if  $\alpha$  and/or $ T \rightarrow  0$ (see below), which is in contrast to the known result. This difficulty can be solved by observing that because in the $zeta$ regularization method the renormalization condition is not an a priori assumption, we may regard the result (3.8) as being
'renormalized' by $V_{eff} = (l/24 \pi) M^3$  at a and/or $T=0$. Therefore one can obtain a useful result from (3.8) only if T >> 1. This is because when T>>I the temperature- independent terms, which are absent from (3.8), are much less than the temperature-dependent terms, which can be obtained from (3.8).
   We will separately discuss two cases: $\alpha  << 1, T >> 1$ and $\alpha  >> 1, T >> 1$. 

\subsection {$\alpha << 1$, while $T >>1$}

   When $\alpha  << 1$ then $Z_1$, and $Z_2$  defined in (3.7) are very large. Using the properties $K_{3/2}(Z) \rightarrow \sqrt \pi Z^{-1/2} e^{-Z}$ as $ Z >> 1$, we see that $V_{eff}$  in (3.8) is dominated by the summation terms of  $n= \pm1$. Therefore we have the relation

$$\sum_{n,l}{}^{'}[2Z_1^{\nu-3/2} K_{3/2-\nu}(2Z) - Z_2^{\nu-3/2} K_{3/2-\nu}(2Z)] $$
$$ = -\sum_{l=1}{}^{'}4\left(\pi [(\alpha/4 +M^2)(l^2/4\pi^2 T^2 +1/\alpha^2)]^{1/2}\right)^{-3/2}$$
$$\times K_{3/2}\left(2\pi [(\alpha/4 +M^2)(l^2/4\pi^2 T^2 +1/\alpha^2)]^{1/2}]\right)$$
$$ = -(2/\pi)^{3/2}(\alpha^2/4 +M^2)^{-1}[-\alpha^2 exp[-2\pi(\alpha^2/4+M^2)/\alpha]$$
$$ + \sum_{l=0}(l^2/4\pi^2 T^2 +1/\alpha^2)^{-1} exp[-2\pi [(\alpha^2/4+M^2) (l^2/4\pi^2 T^2 +1/\alpha^2)]^{1/2 ]}, $$
$$    ~~~~~~ if ~~~\alpha <<1.   \eqno{(3.9)}$$
\\
From this relation one sees that $V_{eff} = 0 $ when $\alpha =0 $ as mentioned in the above comment.  When $T >> 1$ the summation over $l$  in (3.9) could be approximated by an integration. After neglecting the T-independent terms the equation of (3.9) becomes
$$ -(2/\pi)^{3/2}(\alpha^2/4 +M^2)^{-1}\int d l (l^2/4\pi^2 T^2 +1/\alpha^2)^{-1} $$
$$ \times exp[-2\pi [(\alpha^2/4+M^2)\alpha (l^2/4\pi^2 T^2 +1/\alpha^2)]^{1/2 }]$$
$$ = -(2/\pi)^{3/2}(\alpha^2/4 +M^2)^{-1} 2\pi\alpha T \int dX {1\over X^2+1} exp[-A(X^2+1)^{1/2}],    ~~~~~ if \alpha <<1,~~ T >>1,  \eqno{(3.10)}$$
$$ A = {2\pi\over \alpha} ({\alpha^2\over 4}+M^2)^{1/2}. \eqno{(3.11)}$$
\\
Substituting the value
$$ \int _0^{\infty} dX {1\over X^2+1} exp[-A(X^2+1)^{1/2}] \approx (\pi/A)^{1/2} e^{-A},  ~~~ if  ~ A>>1 ,  \eqno{(3.12)}$$
which is evaluated in the appendix, into (3.10) and combining the tree level effective potential we obtain the final result

$$ V_{eff} =  m^2+ \xi R - (2\pi)^{-2} \alpha^{3/2} T ({\alpha^2\over 4}+M^2) ^{1/4} exp[-{2 \pi \over \alpha} ({\alpha^2\over 4}+M^2)^{1/2}]. \eqno{(3.13)}$$
\\
Taking the limit $\phi \rightarrow 0$ from the above equation we obtain the effective mass

$$ m_{eff} =  m^2+ \xi R + \lambda (2\pi)^{-2} \alpha^{3/2} T  \left({\pi \over \alpha}|{\alpha^2\over 4} +m^2 + \xi R|^{-1/4} - {1\over4}|{\alpha^2\over 4} +m^2 + \xi R|^{-3/4}\right)$$
$$\times exp\left(-{2\pi \over \alpha}|{\alpha^2\over 4} +m^2 + \xi R|^{1/2}\right). \eqno{(3.14)}$$
\\
From the above equation we can find the critical temperature by setting $m_{eff}$see that the symmetry will be restored at high temperature.

\subsection {$\alpha  >> 1$ and $T>> 1$}

When $\alpha  >> 1$ and $T>> 1$ then $Z_1$ and $Z_2$ defined in (3.7) are very small if n and l are not too large. In this case the summations over n and I in (3.8) could be approximated by the integrations. Also, because $K_{2/3}(Z) \rightarrow  Z^{-3/2} - {1\over 2}Z^{1/3}$ we see that the main contributions in (3.8) are those coming from the summations over small n and l. Therefore we have the relation

$$ \sum_{n,l}{}^{'}[2Z_1^{\nu-3/2} K_{3/2-\nu}(2Z) - Z_2^{\nu-3/2} K_{3/2-\nu}(2Z)]$$
$$= {2T\alpha \over \pi^2}\left({\alpha^2 \over 4} + M^2\right)^{-3/2} S_1 +  \sqrt 2 T \alpha \left({\alpha^2 \over 4} + M^2\right)^{-1/2} S_2   \eqno{(3.15)}$$
\\
where

$$ S_1(T,\alpha) = \int_{X_0}^{\tilde X} dX \int_{2Y_0}^{\tilde Y} dY (X^2+Y^2)^{-3/2} - \int_{X_0}^{\tilde X} dX \int_{Y_0}^{\tilde Y} dY (X^2+Y^2)^{-3/2}                      $$
$$ S_2(T,\alpha) = \int_{X_0}^{\tilde X} dX \int_{2Y_0}^{\tilde Y} dY (X^2+Y^2)^{-1/2} - \int_{X_0}^{\tilde X} dX \int_{Y_0}^{\tilde Y} dY (X^2+Y^2)^{-1/2}                      \eqno{(3.16)}$$

$$                X_0 = 1/2 T , ~~   Y_0= 1/\alpha      \eqno{(3.17)}$$
\\
while $\tilde X$,  and $\tilde Y$, being independent of T and $\alpha$, stand for the suitable values corresponding to the reasonable cut-off values of n and I. Notice that the factor $(\alpha ^2 /4+ M^2)^{-3/2}$ in the first term of (3.15) will be canceled by the factor $(\alpha ^2/4 + M^2)^{3/2}$ in equation (3.8). Therefore the first term in (3.15) merely contributes a $\phi$-independent term to the effective potential, which at most shifts a value to the effective potential, and we can drop it. Combining the tree level effective potential we finally obtain the result

$$ V_{eff} = [m^2 + \xi R + {1\over 8}\lambda (2\pi)^{-3/2}\alpha T S_2] \phi^2.    \eqno{(3.18)}$$
\\
From the above equation we also see that the symmetry will be restored at high
temperature.  Note that the critical temperature can be easily obtained if  $ T>>\alpha >>1$,  because in this case we may set $X_0 = 0$ and thus $S_2$ is only a function of  $\alpha$.

    Note that in the case $\alpha ^2/4  + m^2 + \xi R \rightarrow 0$ then (3.14) seems to give a divergent $m_{eff}$, which is intuitively an impossible occurrence. The reason is that if $\alpha ^2/4  + m^2 + \xi R \rightarrow 0$ then $Z_1$ and $Z_2\rightarrow 0$,  (if $\phi$  also approach zero), this corresponds to the case of $\alpha, T >> 1$.  This means that for the models with several values of' $\alpha$, $\xi$ and $m^2$  (note that $R = -\alpha^2$) which makes $\alpha ^2/4  + m^2 + \xi R \rightarrow 0$ we shall have the effective potential described by (3.18).

\section { Discussions}
After the new scenario of' the inflationary Universe [2, 3] was proposed, many authors have been attracted to investigate the quantum-field effects on the cosmological phase transition [7, 8]. The problem with the anisotropic effect on the Coleman-Weinberg symmetry breaking has been studied [10-12]. We have also in a recent publication [ 13] studied the inhomogeneous effect on the Coleman-Weinberg symmetry breaking.

    Recently, Chimento et al [ 14] calculated the 1-loop effective potential for the scalar field in a Godel spacetime, and found that whether the symmetry was restored or not would depend only upon the curvature coupling constant $\xi$, irrespective of the magnitude of the rotation. 

    As the early Universe is also characterized by non-zero temperature [17, 18] it is
worthwhile to study the temperature effect on the cosmological phase transition in a rotating spacetime. In this paper we have calculated the finite temperature l-loop
effective for a $\phi ^4$ theory in the Godel spacetime. From our results one can find the critical temperature for the symmetry restoration, which in general depends on the curvature coupling constant and magnitude of the spacetime rotation. 

    Finally we want to make a remark. As Godel spacetime is not very physical our
calculation is not intimately related to the direct observation. However, the investigation in this paper will be (at least, we hope) helpful in the future analysis of the finite-temperature cosmological phase transition in the more general and more realistic (physical) spacetimes.  

\vspace{3cm}
{\bf  \Large Appendix}

Calculation of the integration in (3.12)

 $$\int _0^\infty dX(X^2+l)^{-1/2} exp[-A(X^2+l)^{1/2}]= \int_1^\infty  dY (Y^2-1)^{-1/2} Y^{-1} e^{-Ay}$$
$$= e^{-A}\int_0^\infty dZ (Z+1)^{-1}[Z(Z+2)]^{-1/2} e^{-ZA}$$
$$= e^{-A}\int_0^\infty dZ [Z(Z+2)]^{-1/2} e^{-ZA}$$
$$- e^{-A}\int_0^\infty dZ Z^{1/2} (Z+1)^{-1}[Z(Z+2)]^{-1/2} e^{-ZA}$$
\\
The existence of an exponential factor $e^{-AZ}$ in the integrand tells us that the dominant contribution of the integration comes from the neighbourhood around Z = 0 when $A >> 1$ (i.e. $\alpha << 1$, see the relation (3.11)). Because the first integral has a factor $Z^{-1/2}$ while the second integral has a factor $Z^{1/2}$, the first integral must be larger than the first one. The first integral can be found exactly [26] and we have the result 

 $$\int _0^\infty dX(X^2+l)^{-1/2} exp[-A(X^2+l)^{1/2}] = K_0(A)$$
\\
where K is the modified Bessel function. Taking the expansion of $K_0$ for $A >> 1$  the equation (3.12) is obtained.

\newpage
\begin{enumerate}
\item Guth A 1981 Phys. Rev. D 23 347
\item  Linde A D 1982 Phys. Lett. 108B 489
\item Albrecht A and Steinhardt P J 1982 Phy, Rev. Lett. 48 1220
\item Langacker P 1981 Phys. Rep. 72 185
\item Coleman S and Weinberg E 1973 Phys. Rev. D 7 I888
\item  Birrell N D and Davies P C W 1982 Quantum Fields in Curved Space (Cambridge: Cambridge University Press)
\item Hu B L 1983 Proc loth Int. Conf. on General Relativity and Gravitation ed B Bertotti, F deFetice and A Pascolini (Roma: Consiglio Nazionale Delle Ricerche) p 1086
\item Abbott L F 1981 Nucl. Phys. B 185 233\\
      Allen B 1983 Nucf, Phys. B 226 228; 1983 Ann. Phys., NY 161 152\\
      Vilenkin A t983 Nuct. Phys. B 226 504\\
      Ford H and Vilenkin 1982 Phys. Rev. D 26 t231
\item  Hartle J B and Hu B L 1980 Phys. Rev. D 21 2756\\
      Hu B L and Parker L 1978 Phys. Rev. D 17 933
\item Critchley R and Dowker J S 1982 J. Phys. A: Math. Gen, 15 157
\item O'Connor D J, Mu B L and Shen T C 1983 Phys. Lett. 130B 31\\
      Shen T C, Hu B L and O'Connor D J 1985 Phys. Ret,. D 31 240I\\
      Hu B L and Shen T C 1986 Phys. Lett. 178B 373\\
      Shen T C and Sobczyk J 1987 Phys. Rev. D 36 397
\item Futamase T 1984 Phys. Rev. D 29 2783\\
      Huang Wung-Hong 1990 "Quantum Field Effect on Symmetry Breaking and Restoration in Anisotropic Spacetimes", Phys. Rev. D 42 1287, [gr-qc/0401037]
\item Huang W-H 1991 "Coleman Weinberg Symmetry Breaking in the Early Universe with an Inhomogeneity " Class. Quantum Grav. 8 83
\item  Chimento L P, Jakubi A S and Pullin J 1989 Class. Quantum Grav. 6 L45
\item  Godel K 1949 Rev. Mod. Phys. 21 447
\item Bernard C W 1974 Phys. Rev. D 9 3312\\
      Dolan L and Jackiw R 1974 Phys. Rev. D 9 3320\\
      Weinberg S 1974 Phys. Rev. D 9 3357
\item Kennedy 1981 Phys. Rev. D 23 I184\\
      H u B L 1982 Phys. Lett. 108B 19\\
      O'Connor D J, Hu B L and Shen T C 1983 Phys. Lett. 130B 31
\item Hu B L, Critchley R and Stylianopoutos 1987 Phys. Rev. D 35 510
\item Hawking S W and Ellis G F R 1973 The Large Scale Structure of Spacetime (Cambridge: Cambridge University Press)
\item Hawking S W 1977 Commun. Math. Phys. 55 133
\item Mashhoon B 1978 Phys. Rev. D I1 2679
\item Hiscock W A 1978 Phys. Rev. D 17 1497
\item Weinberg S 1972 Gravitation and Cosmology (New York: Wiley) section 4.4
\item Rubin M A and Roth B D 1983 Phys. Lett. 127B 55
\item Birmingham D and Sen S 1985 Ann. Phys., NY 161 121
\item  Gradsbteyn I S and Ryzhik I M I980 Table oftntegrafs, Series, and Products (New York: Academic)
\end{enumerate}
\end{document}